\documentclass{ims9x6}

\usepackage{mathtools}
\usepackage{siunitx}[=v2]
\DeclareSIUnit\hartree{hartree}
\DeclareSIUnit\bohr{bohr}
\DeclareSIUnit\calorie{cal}

\usepackage{rotating}

\makeatletter
\renewcommand\appendix{
  \vspace{10pt plus 2pt minus 2pt}%
  \section*{Appendices}%
    \refstepcounter{appendix}%
    \setcounter{section}{0}%
        \setcounter{lemma}{0}
        \setcounter{theorem}{0}
    \setcounter{definition}{0}
        \setcounter{corollary}{0}
    \setcounter{equation}{0}
    \@addtoreset{equation}{section}
\renewcommand\thesection{\Alph{section}}%
\renewcommand\thesubsection{\Alph{section}.\arabic{subsection}}%
\renewcommand\theequation{\Alph{section}.\arabic{equation}}
\def\@seccntformat##1{{\upshape{\csname the##1\endcsname}.}\hskip .5em}
}%
\makeatother

\newcommand{\R}{\mathbb{R}}  

\DeclareMathAlphabet{\vecfont}{OT1}{cmr}{bx}{it}

\newcommand{\bfr}{\vecfont{r}}

\newcommand{\bfR}{\vecfont{R}}

\DeclarePairedDelimiter\bra{\langle}{\rvert}
\DeclarePairedDelimiter\ket{\lvert}{\rangle}
\DeclarePairedDelimiterX\mel[3]{\langle}{\rangle}
  { #1 \delimsize\vert \mathopen{}#2 \delimsize\vert \mathopen{}#3 }
\DeclareMathOperator{\erf}{erf}
\DeclareMathOperator*{\argmax}{arg\,max}
\DeclareMathOperator*{\argmin}{arg\,min}
\DeclarePairedDelimiterX\norm[1]{\lVert}{\rVert}{#1}
\DeclarePairedDelimiterX\abs[1]{\lvert}{\rvert}{#1}

\newcommand{\hminus}{H\textsuperscript{--}}
\newcommand{\hplus}{H\textsuperscript{+}}
\newcommand{\htwo}{H\textsubscript{2}}

\newcommand*{\Eqref}[1]{Eq.~\eqref{#1}}
\newcommand*{\D}{\mathrm{d}} 
\newcommand*{\grad}{\boldsymbol{\nabla}}

\makeatletter
\renewcommand{\ps@plain}{%
  \renewcommand{\@oddhead}{\hfil\footnotesize%
    \raisebox{30pt}[0pt][0pt]{\parbox{300pt}{\centering%
      A contribution to the Proceedings of the\\{}%
      Workshop on Density Functionals for Many-Particle Systems\\{}
      2--29 September 2019, Singapore}}\hfil}%
  \renewcommand{\@evenhead}{\@oddhead}%
  \renewcommand{\@oddfoot}{\hfil\footnotesize%
        \raisebox{-8pt}[0pt][0pt]{\thepage}\hfil}%
  \renewcommand{\@evenfoot}{\@oddfoot}%
}
\makeatother
\renewcommand{\trimmarks}{\relax}

\begin{document}

\chapter{FLEIM:\\ A stable, accurate and robust extrapolation method at
  infinity for  computing the ground state of electronic Hamiltonians}

\author{\'Etienne Polack}
\address{Universit\'e
  Bourgogne Franche-Comt\'e and CNRS,\\
  Laboratoire de Math\'ematiques de Besan\c con, F-25030 Besan\c con,
  France\\[0.5ex]
  etienne.polack@math.cnrs.fr}
\author{Yvon Maday}
\address{Sorbonne Universit\'e, CNRS, Universit\'e de Paris,\\
  Laboratoire Jacques-Louis Lions (LJLL), F-75005 Paris,\\
  and Institut Universitaire de France\\[0.5ex]
  yvon.maday@sorbonne-universite.fr}
\author{Andreas Savin}
\address{CNRS and Sorbonne Universit\'e,\\
  Laboratoire de Chimie Th\'eorique (LCT),
  F-75005 Paris, France\\[0.5ex]
  andreas.savin@lct.jussieu.fr}

\markboth{\'Etienne~Polack, Yvon~Maday, and Andreas~Savin}%
         {FLEIM for electronic Hamiltonians}

\begin{abstract}
  The Kohn--Sham method uses a single model system, and corrects it by a density functional
  the exact user friendly expression of which is not known and is replaced by an
  approximated, usable, model. We propose to use instead more than one model system, and use
  a greedy extrapolation method to correct the results of the model systems. Evidently,
  there is a higher price to pay for it. However, there are also gains: within the same
  paradigm, e.g., excited states and physical properties can be obtained.
\end{abstract}


\vspace*{12pt}

\section{Introduction}
\subsection{Motivation}
Density functional theory (DFT) has a weak point: its approximations (DFAs).
First, the Hohenberg--Kohn theorem tells us that there is a density functional for electronic
systems, $F[\rho]$, that is universal (that is, independent of the potential of the nuclei),
but does not give us a hint on how systematic approximations can be constructed.
In practice, models are produced to be fast in computations, typically by transferring
properties from other systems, like the uniform electron gas.
Second, the most successful approximations are using the Kohn--Sham method (introducing a
fermionic wave function) that decomposes $F[\rho]$ into the kinetic energy,
the Hartree energy and an
exchange-correlation energy contribution although the question of how and what part of
$F[\rho]$ should be approximated is, in principle, open.

In the present contribution we totally change the paradigm in the following way still led by the
issue of universality.
Let us start with a physical consideration.
When electrons are close, the Coulomb repulsion is so strong that some of its features
dominate over the effect of the external potential.
This is also reflected mathematically in the short-range behavior of the wave function, as
present in the Kato cusp
condition~\cite{Kat-CPAM-57,FouHofHofOst-CMP-09,Yse-10,FlaFlaSch-20}, and in higher-order
terms~\cite{RasChi-JCP-96a,KurNakNak-AQC-16}.
We further note that approximating numerically the short-range part of the wave function
needs special care, due to the singularity of the Coulomb interaction when the electrons are
close.

 The considerations above and the independence of the interaction between electrons from
that between them and the external potential provides a basis for constructing
approximations.
Thus, we propose to solve accurately a Schr\"odinger equation with a Hamiltonian that is
modified to eliminate the short-range part of the interaction between the electrons which is
one of the difficult parts in the numerical simulations.
The way to do it is not unique, and we try to turn this to our advantage: we use several
models, and from them we try to extrapolate to the physical system~\cite{Sav-JCP-11}.
In other words, we follow an ``adiabatic connection'' (see~\cite{HarJon-JPF-74}),
without ever constructing a density functional. This new paradigm thus explores the
possibility to replace the use of DFAs by mathematically controlled approximations: we make
density functional theory ``without density functionals.''

\enlargethispage{0.8\baselineskip}

Our approach has introduced an additional difficulty nonexistent in the Kohn--Sham method:
the long-range part of the interaction has to be treated accurately, and not only its
electrostatic component.
One may ask whether this additional effort is justified, and whether one gains anything with respect
to a calculation where the physical (Coulomb) interaction is used.
For a single calculation, the gain is due to the lack of singularity in the interaction
expressed by a weak interaction potential allowing for
simplified treatments, such as perturbation theory.
However, as the extrapolation to the physical system needs more than one point, it is
essential that the number of points stays very small, and the interaction weak.

\subsection{Objective and structure of the paper}
We first choose, in Sec.\ref{sec:2.1}, a family of model (parameter dependent) Hamiltonians that are more flexible
than using only the Kohn--Sham (noninteracting) Hamiltonian.\footnote{Note however that this
is at the prize of working in $\R^{3N}$ instead of $\R^3$, and thus requiring accurate
many-body, e.g., configuration interaction calculations.}
This is followed by a description of how universality is introduced, namely by analyzing how
a nonsingular interaction approaches the Coulomb one, and not by transfer from other
systems, as usually done in DFAs.
The physical system of interest is one among the parameter dependent models corresponding to
some precise value of the parameter;
in Sec.~\ref{sec:2.2} its solution is extrapolated from the solutions to the
models for other values of the parameter, expected that these solutions are more simple to
be approximated.
This extrapolation is efficiently handled  in the general framework
of the model reduction methods and more precisely  referring to  a variation of
the \emph{Empirical Interpolation Method}~\cite{barrault_empirical_2004}.

We believe that such an approach can not only discuss what DFAs are really doing, but can
evolve to being used in applications.
Some argument supporting this statement is given.
However,  in this paper numerical examples (gathered in Sec.~\ref{sec:3}) are
only presented for two-electron 
systems that are numerically (and sometimes even analytically) easily accessible: the
harmonium, the hydrogen anion, \hminus, and the hydrogen molecule, \htwo{} in the ground
state, at the equilibrium distance.

As we do not use the Hohenberg--Kohn theorem, the technique can be applied without
modification also to excited states.
We provide in Sec.~\ref{sec:3.5}, as an example, the first excited state of
the same symmetry as the ground state.

Some conclusions and perspectives are presented in Sec.~\ref{sec:4}.
Finally, in order to facilitate reading the manuscript, various details are given in
Appendices \ref{app:DFT}--\ref{app:change-to-cm-12}
that follow Sec.~\ref{sec:4}.

\section{Approach}
\subsection{The model Schr\"odinger equation}\label{sec:2.1}
We study a family of Schr\"odinger equations,
\begin{equation}
  \label{eq:schroedinger}
  H(\mu) \Psi(\mu) = E(\mu) \Psi(\mu),
\end{equation}
where $\mu$ is some nonnegative parameter.
More precisely, in this paper, we use
\begin{equation}
  \label{eq:H}
   H(\mu) = T + V + W(\mu),
\end{equation}
where $T$ is the operator for the kinetic energy, $V$ is the external potential (in
particular that of the interaction between nuclei and electrons) and $W(\mu)$ represents the
interaction between electrons. Although not required by the general theory, in this paper we
introduce the dependence on $\mu$ only by modifying the interaction between electrons,
\begin{equation}
  \label{eq:W}
  W(\mu) = \sum_{i<j} w(r_{ij},\mu),
\end{equation}
choosing
\begin{equation}
  \label{eq:werf}
  w(r_{ij},\mu) = \frac{\erf(\mu r_{ij})}{r_{ij}}
\end{equation}
where $r_{ij}=\abs{\bfr_i - \bfr_j}$ is the distance between electron $i$ (at position
$\bfr_i$) and electron $j$ (at position $\bfr_j$). Finally, the external potential $V$ is
written like
\begin{equation}
  V = \sum_{i=1}^N v(\bfr_i).
\end{equation}
where $v$ is the local one particle operator. Note that the $N$-particle operators are
denoted by upper case letters, while the one-particle operators are denoted by lower case
letters.

Note also that for $\mu=0$ we have a trivial noninteracting system, while for $\mu=\infty$
we recover the Coulomb system.
The operator $w$ is long-ranged: as $\mu$ increases, the Coulomb interaction $1/r_{12}$
starts being recovered from large distances.
The first reason for this choice is that, as mentioned above, we expect a universal
character for short range (this is related to the difficulty of common DFAs to correctly
describe long-range contributions, cf.\ Appendix~\ref{app:DFT}).
The second reason is that the solution of~\Eqref{eq:schroedinger} is converging more rapidly
with (conventional) basis set size when the interaction has no singularity at $r_{12}=0$.

In principle, introducing a dependence of the one-particle operators ($T$ and $V$) on $\mu$
makes the formulas a bit more clumsy, but does not introduce important difficulties in its
application.
Using such a dependence might improve the results, but it is not discussed in this contribution.
In the following, in order to simplify notation, we drop the argument $\mu$, when
$\mu=\infty$, e.g., $E=E(\mu=\infty)$.

\subsection{The correction to the model}\label{sec:2.2}
\subsubsection{Using a basis set}\label{Using_a_basis_set}
Of course, solving the Schr\"odinger equation for the model, \Eqref{eq:schroedinger} with
finite $\mu$s, does not provide the desired solution, i.e., the one that is obtained for
$\mu=\infty$.
We thus need to estimate the difference in eigenvalues:
\begin{equation}
  \label{eq:ebar}
  \bar{E}(\mu) = E - E(\mu).
\end{equation}
Since $\bar{E}(\mu)$ tends to zero at infinity, the idea is to first expand this difference
$\bar{E}(\mu)$ in a basis (of functions that tend to zero at infinity), retaining $M$ terms,
 \begin{equation}
   \label{eq:e-basis}
   \bar{E}(\mu) \approx \bar{E}_M(\mu) = \sum_{j=1}^M c_j \chi^{\ }_j(\mu),
\end{equation}
leading to
\begin{equation*}
  E(\mu) \approx E - \sum_{j=1}^M c_j \chi^{\ }_j(\mu),
\end{equation*}
or, more precisely since $E$ is not known, we replace it by an approximation denoted as
$E_M$,
\begin{equation}
  \label{eq:e-mu-basis}
  E(\mu) \approx E_M - \sum_{j=1}^M c_j \chi^{\ }_j(\mu).
\end{equation}
The idea then proceeds by determining the unknown $E_M$ values and the coefficients $c_i$ from
$M+1$ values of $E(\mu_m)$, for~$m=0, \dotsc, M$ for an appropriate choice of the parameter
values $\mu_m$.
Finally, taking into account that the functions $\chi^{\ }_j$ tend to zero at infinity, the
proposed approximation for $E$ is $E_M$.
Of course, this extrapolation approach often fails if care is not enough taken in the choice
of the functions $\chi^{\ }_j, 1\le j\le M$, and the values $\mu_m$, for~$m=0, \dotsc, M$.

First, one has to decide about their form.
Second, one has to find a way to keep $M$ as small as possible to reduce computational cost
while preserving a good accuracy.

\subsubsection{Approaching the Coulomb interaction}
As recalled above, we derive from the leading term of the Coulomb interaction between the
electrons that, to leading order, the solutions of differential equations are determined at
short range by the singularities.
The interaction $w$ in \Eqref{eq:werf} has no singularity at $r_{12}=0$, for any finite $\mu$.
However, as the parameter $\mu$ increases, $w(\cdot, \mu)$ approaches the singular Coulomb
potential.

In order to see how this limit is approached, let us perturb the exact solution.
To first order, the perturbation correction to the energy is given by
\begin{equation}
  \label{eq:pert}
  \bar{E}(\mu) = \mel*{\Psi}{\bigl(W - W(\mu)\bigr)}{\Psi},
  \quad \text{for \( \mu \rightarrow \infty \)}.
\end{equation}
By changing the integration variables $\bfr_i$ to $\mu \bfr_i$ we see that
\begin{equation}
  \label{eq:approach}
  \bar{E}(\mu) \propto \mu^{-2}
  \quad \text{as \( \mu \rightarrow \infty \)},
\end{equation}
providing a leading behavior that we want the basis functions $\chi^{\ }_i$ to reflect.
It is possible to continue this analysis for higher order terms.
In fact, the next term (in $\mu^{-3}$) has a coefficient proportional to that of $\mu^{-2}$,
the proportionality coefficient being determined by the nature of the Coulomb
singularity~\cite{GorSav-PRA-06}.

\subsubsection{Choice of the basis functions}
In the main part of this contribution we use a simple ansatz,
\begin{equation}
  \label{eq:basis}
  \tilde \chi_j^{\ }(\mu)=1- j \; \mu (1+ j^2 \; \mu^2)^{-1/2},
  \quad j=1, \dotsc, M,
\end{equation}
that respects indeed the condition of \Eqref{eq:approach}.
The motivation for this specific choice, that is arbitrary to a certain degree, as well as
some results obtained with other choices of basis functions, is given in
Appendix~\ref{app:basis}.

\begin{figure}[htb]
  \centering
  \includegraphics{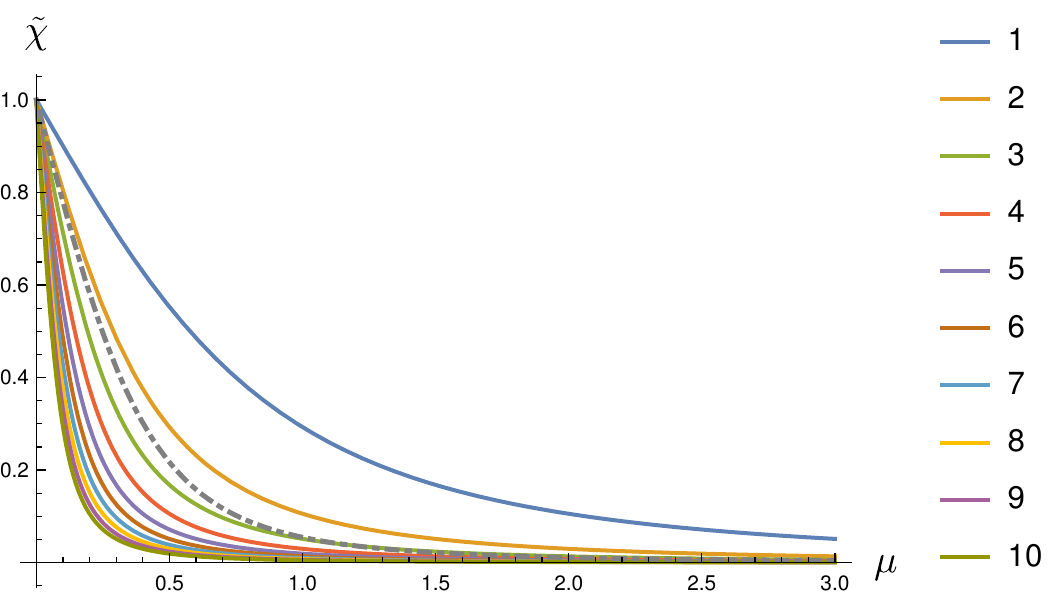}
  \caption{Basis functions $\tilde \chi^{\ }_j$ of \Eqref{eq:basis}, continuous
    curves with the color corresponding
    to $j$; and an (unknown) function to be approximated by linear combination
    on this basis (dot-dashed, gray).
    The unknown function in this figure is proportional to $\bar{E}(\mu)$
    of harmonium.\label{fig:basis}}
\end{figure}

The first functions of this basis set are presented in Fig.~\ref{fig:basis}, together with
an example of a function it has to approximate. It illustrates that the function we want to
describe is between basis function $\tilde\chi_2^{\ }$ --- for small $\mu$ ---
and basis function $\tilde\chi_3^{\ }$ ---for
large $\mu$.
However, a simple linear combination between these (only) two surrounding
basis functions from the family in
\Eqref{eq:basis} does not improve much the accuracy, but of course, more (and more
appropriate) functions in the family can (and should) be called.

\subsubsection{Reducing the basis set}
Using a large set of $\chi_j^{\ }$ (a large $M$) can rapidly become computationally prohibitive
(because it requires a large number of evaluations of $E(\mu_m)$, for~$m=0,\dots, M$) and
numerically unstable (because it is classically much more difficult to stabilize
extrapolation than interpolation).
In order to reduce their number and increase the stability of the extrapolation, we use a
greedy (iterative) method, as in the \emph{Empirical Interpolation Method} (EIM) leading to
proper choices of $\mu_m$, for~$m=0,\dots, M$ known as ``magic points.''

In the $K$th iteration of EIM, one starts from a set of $K-1$ basis functions (for us,
$\tilde \chi^{\ }_j$) and $K-1$ points (for us, $\tilde \mu_j$ belonging to some (discretized) interval, say,
close to zero, to benefit at most of the regularization of the $\erf$ function). One then
chooses the $K$th function $\tilde \chi^{\ }_{K-1}$ (among the remaining $M-K$ basis
functions) as being the one that is most poorly approximated by the current interpolation
(based on the $K-1$ basis functions and the $K-1$ points) in a sense dedicated to the final
goal we want to achieve (that can be uniform error, error on some part of the domain, or
even at some value) and the $K$th point $\tilde \mu_{K-1}$ that, in the admissible set,
brings the more information.
In this contribution, as we are only interested at extrapolating the value of~\( \mu \) at
 infinity, so we chose the error as the absolute value of difference between the $K$th
basis function and its interpolant at infinity as the final goal we want to achieve.

Note that the procedure selecting the next point and function does not make any use of the
function to be approximated (here $\bar{E}$).
It is thus a cheap step compared with the calculation of $E(\tilde \mu_m)$ on the system of
interest.

To improve the results for the extrapolation, we have modified the EIM algorithm into what
we call the \emph{Forward Looking} EIM (FLEIM).
While EIM tries to get the maximal improvement through a sequential choice of, first the new
basis function, then the new point of interpolation, FLEIM tries to get the best pair for
improvement in the selected goal.
The method is explained in more detail in Appendix~\ref{app:EIM}.
In what follows, we present the results of FLEIM as they are better and more stable than
those of EIM, as is illustrated in App~\ref{app:details-calc1}.

\subsection{Computing other physical properties}
FLEIM can be used to approximate other physical properties, i.e., correct expectation values
of operators $A \ne H$ obtained with the model wave functions, $\Psi(\mu)$,
\begin{equation}
A(\mu) = \bra{\Psi(\mu)} A \ket{\Psi(\mu)}.
\end{equation}
This can be seen immediately by noting that the derivation in
Sec.~\ref{Using_a_basis_set} is not specific for correcting $E(\mu)$, but can
also be applied to $A(\mu)$.

For the choice of the basis functions, we point out that properties are obtained by
perturbing the Hamiltonian with the appropriate operator, say, $A$,
\begin{equation}
H \rightarrow H(\lambda) = H + \lambda A.
\end{equation}
The expectation value of $A$ can be obtained as the derivative of $E(\lambda)$ w.r.t.
$\lambda$, at $\lambda = 0$. Of course, this procedure can be applied to model Hamiltonians,
yielding $E(\lambda,\mu)$ and
\begin{equation}
\bra{\Psi(\mu)} A \ket{\Psi(\mu)} = \partial_\lambda E(\lambda,\mu)\bigr|_{\lambda=0}
\end{equation}
Thus, in this contribution, we use the same type of basis functions for $A(\mu)$ as for $E(\mu)$;
see the results in Sec.~\ref{sec:expectation_values}.
Note that computing $\bra{\Psi(\mu)} A \ket{\Psi(\mu)}$ is not possible in
DFT, without having a property-specific density functional~\cite{Bau-PRB-83}.

\section{Numerical results}\label{sec:3}
\subsection{Guidelines}
The quality of the corrections using Eqs.~\eqref{eq:e-basis} and~\eqref{eq:basis} is
explored numerically.
Technical details on the calculations are given in Appendix~\ref{app:details-calc}.

The plots show the errors done by the approximations in the estimate of the energy: we
choose a model, $\mu\mapsto E(\mu)$, and let the empirical interpolation method choose which
easier models (with weaker interactions) to extrapolate and get an approximation for
$E=E(\mu=\infty)$.
The plots show the error in the estimate of $E$ made when considering approximations that
use information only for $\tilde \mu_m \le \mu$.
From the plots, we read off how small $\mu$ can be and still have ``reasonable'' accuracy.
In thermochemistry, \si{\kilo\calorie\per\mole} is a commonly considered unit, and is often
considered as ``chemical accuracy.''
For electronic excitations, one often uses \si{\eV} units, and one often indicates it with
one decimal place.
``Chemical accuracy'' is marked in the plots by horizontal dotted lines.
The plots show the errors in the range of
$\pm \SI{0.1}{\eV} \approx \SI{2.3}{\kilo\calorie\per\mole}$.

We consider approximations using up to four points (thus chosen in $[0, \mu]$).
The first point $\mu_0$ is always the value chosen $\mu_0 = \mu$ shown on the $x$-axis of
the plots, and the basis function associated to it is $\tilde \chi^{\ }_0$, the constant function; note that using only this pair $(\chi^{\ }_0, \mu_0)$ corresponds to choosing $E\simeq E(\mu_0)$=
the value provided by the model, i.e., no correction is applied.
When the number of points is increased, further values of $E(\tilde \mu_m)$, chosen by the
algorithm, are used with $\tilde \mu_m < \mu$.

The (maximal) parameter $\mu$ is considered between \num{0}~and~\SI{3}{\per\bohr}.
The model without correction (blue curve) reaches chemical accuracy for $\mu \approx \SI{3}{\per\bohr}$ for
  \hminus{} and harmonium, but only at $\mu \approx \SI{5}{\per\bohr}$ for \htwo{} in its
  ground state.

\subsection{General behavior of errors}
\begin{figure}[p]
  \centering
  \includegraphics[viewport=140 165 450 635,clip=]{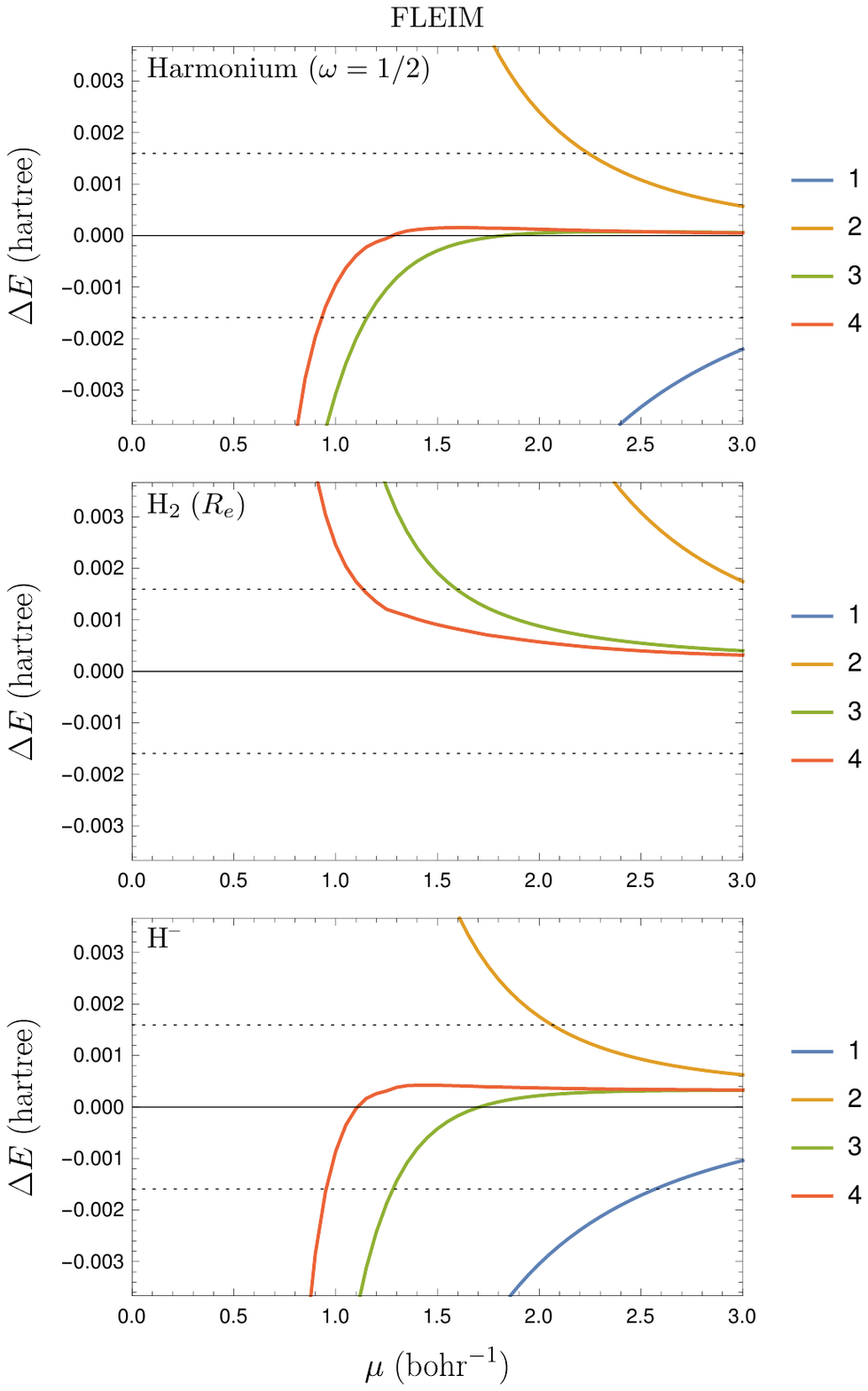}
  \caption{\label{fig:was2-4}%
    Errors for harmonium (top), \htwo{} (middle), and \hminus{} (bottom)
    using FLEIM with one to four points (1: blue curve, 2: brown
    curve, 3: green curve, 4: red curve).
    The abscissa represents the biggest~\( \mu \)
    allowed for use in the FLEIM algorithm.
    The error of the model without correction (blue curve) does not show up 
    in the figure for the \htwo\ molecule because it is larger than the 
    domain covered by the plot.}
\end{figure}

The plots in Fig.~\ref{fig:was2-4} for harmonium, \htwo, and \hminus{} have similar features  and are discussed together.
As the number of points used increases, the smallest value of $\mu$ for which the good
accuracy is reached decreases.
Note that FLEIM produces very small errors for values of $\mu$ larger than 2.
However, with the chosen basis set, the algorithm presented in this contribution has difficulties
correcting the errors for $\mu$ smaller than 1.

\subsection{Possibility of error estimates}
Some tests can be done to estimate the quality of the approximation.
For example, we can compare how the approximations change when increasing the
number of basis functions, $K$, in our approximation and consider
${\abs{E_{K} - E_{K-1}}}$ as an \emph{asymptotically} valid error estimate for
$E_{K-1}$. 
One can notice in the above figures that, when the difference between, say,
the 2- and the 3-point approximation error is larger than ``chemical
accuracy,'' so is the error in the 2-point approximation.

\subsection{Expectation values with FLEIM: $\langle r_{12} ^{\  }\rangle$ and
  $\langle r_{12} ^2\rangle$ for harmonium}\label{sec:expectation_values}
\begin{figure}[t]
  \centering
  \includegraphics[viewport=140 295 450 660,clip=]{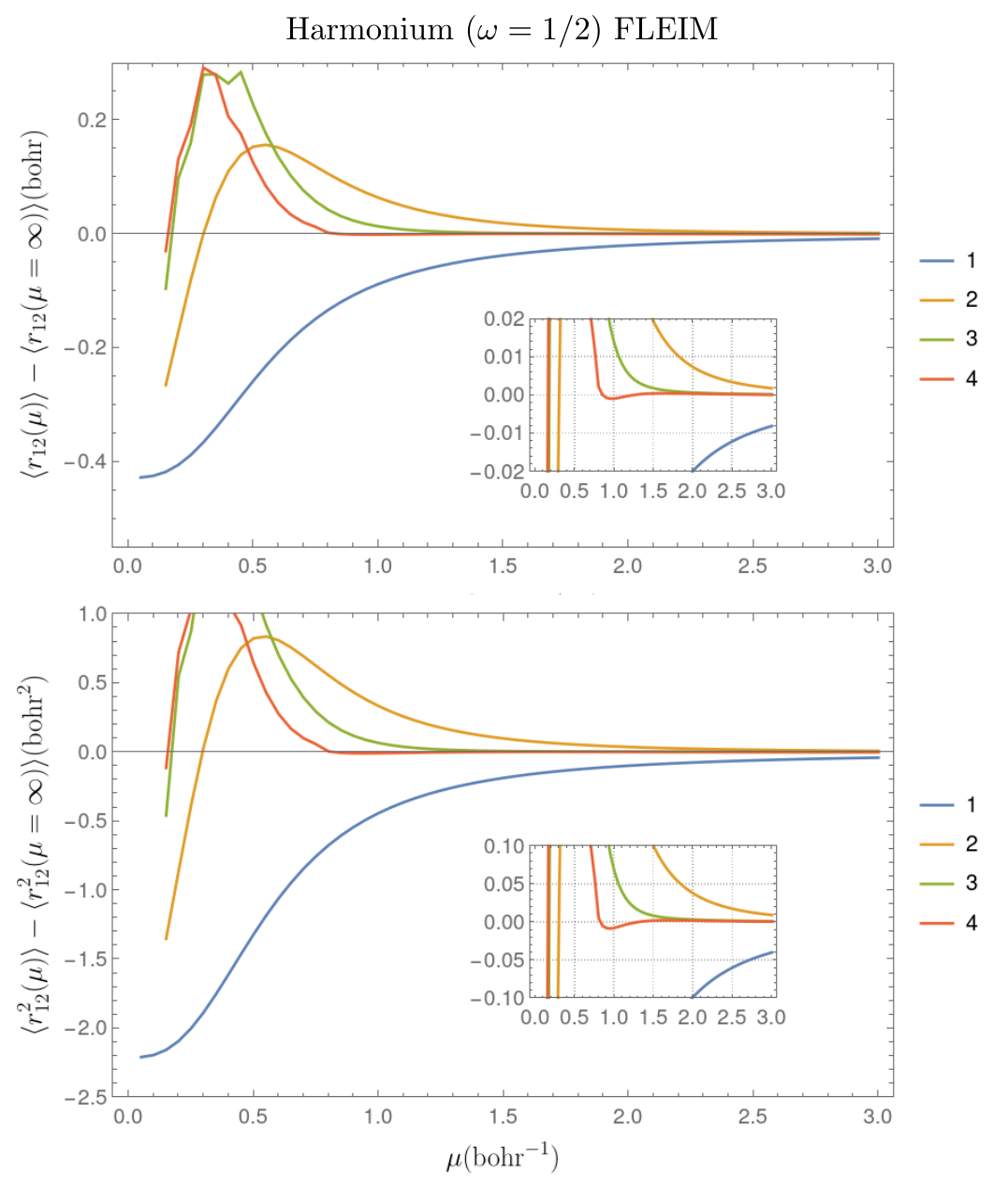}
  \caption{\label{fig:was5-6}%
Errors made for the expectation value of the distance between electrons (top)
and the distance squared (bottom) for harmonium, by using a model wave
function, $\Psi(\mu)$, and after correcting with FLEIM (the different curves
correspond to the number of points used). The insets zooms in.}
\end{figure}

We look at the average distance between the electrons in harmonium.
Figure~\ref{fig:was5-6}(top) shows the error made by using $\Psi(\mu)$ instead
of $\Psi(\mu=\infty)$ in computing the expectation value of $r_{12}$, as well
as the correction that can be achieved with FLEIM, using the same basis set as
above (\ref{eq:basis}). 
The inset in Figure~\ref{fig:was5-6}(top) concentrates on the errors made in
the region that could be considered chemically relevant
(\SI{1}{\pico\meter}~$\approx$~\SI{0.02}{\bohr}).
We note the similarity with the behavior of in correcting $E(\mu)$.

Let us now examine the average square distance between the electrons,
$\langle r_{12}^2 \rangle$, in harmonium.
While for computing the energy we explored correcting the missing short-range part of the
interaction, we now ask whether it is possible to correct the error of using the model wave
function, $\Psi(\mu)$ for the expectation value of an operator that is important at long
range.

For $\omega=1/2$, we know the exact values of the expectation value of $r_{12}^2$ at $\mu=0$
and $\mu=\infty$; they are 6 and $ (42\sqrt{\pi}+64)/(5\sqrt{\pi}+8) \approx 8.21$,
respectively (see, e.g., Ref.~\cite{Kin-96}).
Note the large effect of the model wave function, $\Psi(\mu)$, in computing
$\langle r_{12}^2 \rangle$.
Figure~\ref{fig:was5-6}(bottom) shows the error made by using $\Psi(\mu)$ instead of
$\Psi(\mu=\infty)$ in computing the expectation value of $r_{12}^2$, as well as the effect
of the correction that can be achieved with FLEIM, using the same basis set (\ref{eq:basis})
as above.
We note again the similarity with the behavior of in correcting $E(\mu)$ or the expectation
value of the distance between electrons.

The expectation value $\langle r_{12}^2 \rangle$ also illustrates another aspect: the effect
of a change of the external potential on the energy.
At first sight this may seem surprising, as the external potential is a one-particle
operator, while $r_{12}^2$ is a two-particle operator.
However, changing the one-particle operator also modifies the wave function
and this affects the value of $\langle r_{12}^2 \rangle$.
In the case of harmonium, this can be shown analytically.
Changing $\bfr_1$ and $\bfr_2$ to center-of-mass, $\bfR$, and inter-particle distance,
$\bfr_{12}$, cf.\ Appendix~\ref{app:change-to-cm-12}, allows us to see explicitly that a
modification of $\omega^2$, the parameter that specifies the external potential, affects the
Sch\"odinger equation in $\bfr_{12}$.
It introduces a term proportional to $\omega^2 r_{12}^2 $.
The first order change in the energy when we change the external potential ($\omega^2$) is
thus proportional to $\langle r_{12}^2 \rangle$.
Our results in Fig.~\ref{fig:was5-6}(bottom) show that our conclusions on
model corrections are not modified by small changes in the external potential.
Note that the center-of-mass Schr\"odinger equation also depends on
$\omega^2$, but it is independent of $\mu$ and thus does not affect our
discussion on model correction.

\subsection{Comparison with DFAs}\label{sec:3.5}
Instead of using extrapolation with FLEIM, one can use DFAs.
While up to now the external potential did not change with $\mu$, in DFA
calculations a one particle potential that depends on $\mu$ is added in order to correct the density.

We consider here two DFAs, the local density approximation,
LDA~\cite{Savin-1996,Paziani_Moroni_Gori-Giorgi_Bachelet_2006}, and one that reproduces that
of Perdew, Burke and Ernzerhof
(PBE)~\cite{Goll_Werner_Stoll_2005,Goll_Leininger_Manby_Mitrushchenkov_Werner_Stoll_2008} at
$\mu=0$.
Both approximations are modified to be $\mu$-dependent.
In particular, they vanish at $\mu=\infty$.

\begin{figure}[t]
  \centering
  \includegraphics[viewport=140 265 450 655,clip=]{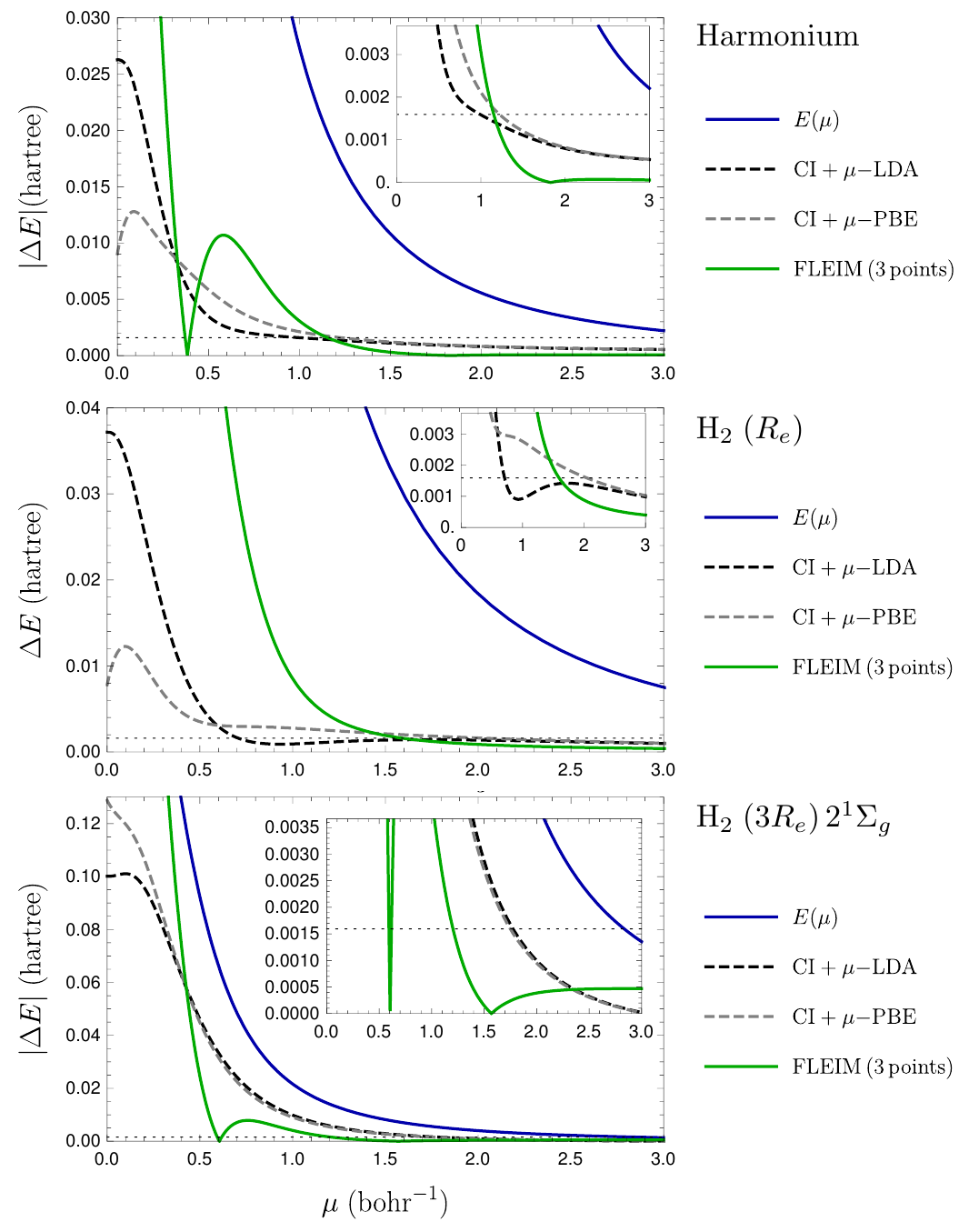}  
  \caption{\label{fig:was7-9}%
Absolute errors for the harmonium molecule at equilibrium distance (top), for
the \htwo{} molecule at equilibrium distance (middle), and in the first
excited state of the same symmetry as the ground state
(bottom):
a $\mu$-dependent LDA (black dashed curve), combined with the a
$\mu$-dependent Perdew--Burke--Ernzerhof approximation (PBE, gray dashed
curve), FLEIM (3 points) (green curve), combined with a $\mu$-dependent local
density approximation.
The abscissa represents the biggest~\( \mu \) allowed for use in the FLEIM
algorithm.
The insets zoom into the regions of small errors, the dotted line corresponding
to the value of ``chemical accuracy.''
Note the different ranges for ${\Delta E}$.} 
\end{figure}

As shown in Fig.~\ref{fig:was7-9}(top) for harmonium, DFAs are clearly much
better at small $\mu$.
However, they are not good enough.
The figure suggests the range of $\mu$ for which DFAs are within chemical
accuracy is similar to that obtained with the 3-point FLEIM\@.
This is confirmed when comparing the results with DFAs and for the \htwo; see
Fig.~\ref{fig:was7-9}(middle).
Note that with FLEIM the errors at large $\mu$ are smaller.

Note also that the curves obtained with extrapolation are significantly \emph{flatter} at
large $\mu$ than those obtained with DFAs.
This should not be surprising: DFAs transfer the large $\mu$ behavior, while extrapolation
extracts it from information available for the system under study.

Furthermore, using ground-state DFAs for excited states does not only pose a problem of
principle (questions its validity, as the Hohenberg--Kohn theorem is proven for the ground
state), but can also show a deterioration of quality.
However, there is no question of principle from the perspective of this contribution (of using a
model and correcting it by extrapolation).
Also, the error in the excited state seems comparable to that in the ground state, as seen
in the example of the \htwo{} molecule, in the first excited state of the same symmetry as
the ground state; see Fig.~\ref{fig:was7-9}(bottom).

\section{Conclusion and perspectives}\label{sec:4}
In this contribution we have illustrated with a few models how to simplify the
Hamiltonians by smoothly getting rid of the singularities in the system and
thus have more numerically tractable problems.
This simplification is obtained by introducing a parameter that, when it is
equal to infinity, it corresponds to the original, plain Hamiltonian.
After numerically solving a few simplified problems, the solution of interest
is obtained by extrapolation.

We present a new (in the field) method for extrapolating the quantities of
interest from few finite values (hence easy to solve) of the parameter by a
technique borrowed from reduced basis paradigm: the empirical interpolation
method. 
In contrast to DFAs, no parameters are fitted, no transfer from different
system are made: only extrapolation is used.
Note that in contrast to DFAs, improvement can be envisaged by either adding
further points or using more appropriate basis functions and error estimates 
are asymptotically accessible.

\appendix

\section{On density functional approximations}\label{app:DFT}

In DFT, the existence of a universal functional of the density,
$F[\rho]$, i.e., the Hohenberg--Kohn theorem~\cite{HohKoh-PR-64}, is rigorously
proven~\cite{Lev-PNAS-79,Lie-IJQC-83}.
However, obtaining accurately the value of $F$ for a given density
$\rho(\bfr)$, while possible (see, e.g., Ref.~\cite{ColSav-JCP-99}), is
exceedingly time-consuming. 
Computationally convenient DFAs exploit the knowledge of
the density around a given point $\bfr$ in space.
Typically,
\begin{equation}
  \label{eq:dfas}
  F[\rho] \approx \int_{\R^3} f \left(
    \rho(\bfr), \abs{ \grad\rho(\bfr) }, \dotsc \right)\,
  \D \bfr.
\end{equation}

The limitation of such an approach can be seen for a simple density
functional, the Hartree term of the energy,\footnote{The volume elements
  $\D\bfr_1^{\ }\,\D\bfr_2^{\ }$ are omitted when the context is unambiguous.}
\begin{equation}
  \label{eq:e-hartree}
  E_H[\rho] = \int_{\R^6}
  \frac{\rho(\bfr_1) \rho(\bfr_2)}{\abs{\bfr_1-\bfr_2}}
\end{equation}
when ${\rho(\bfr) = \rho_A(\bfr) + \rho_B(\bfr)}$ and ${\rho_A(\bfr)
  \rho_B(\bfr) \approx 0}$ 
(i.e., $\rho_A$ and $\rho_B$ are spatially separated; their overlap decreases
much faster than $1/|\bfr_1-\bfr_2|$).
Hohenberg and Kohn recognized the difficulty of approximating $E_H$ by
expressions of the type given in~\Eqref{eq:dfas} and suggested separating it
from $F[\rho]$. 
However, this does not fundamentally solve the problem, as one can immediately
see in 
one-electron systems, where $E_H$ has to be canceled by another term commonly expressed in
DFAs by an ansatz of the forms given in~\Eqref{eq:dfas}.
Note, that the problem would not exist for interactions that are not Coulomb (long-ranged),
but short-ranged.
For instance, if the interaction is Dirac's $\delta(\bfr_1-\bfr_2)$ function, $E_H$ becomes
exactly of the form of~\Eqref{eq:dfas}.
For other short-range interactions one can approach such a form by using Taylor expansions.
In recent years it has become popular to compensate for the limitation of the ansatz
in~\Eqref{eq:dfas} by adding ``empirical'' energy corrections to describe long-range
effects.

Another problem is that the antisymmetry of the electronic wave function is hidden in
$F[\rho]$.
As the most important effect of the antisymmetry is the Pauli repulsion, Kohn and
Sham~\cite{KohSha-PR-65} proposed to consider the variational principle for a model in which
particles do not interact.
However, DFAs following the pattern of~\Eqref{eq:dfas} are still in use.
In fact, further separating terms from $F[\rho]$ may even lead (for degenerate cases) to the
question whether the limit of a noninteracting system is well-defined (see, e.g., Sec.~3.5 in~\cite{Savin-1996}).

\section{Basis functions}\label{app:basis}
In order to get an idea how the leading term of the correction behaves, we
consider the missing part of the Hartree term,
\begin{equation}
  \mel*{\Psi(\mu)}{\bigl(W-W(\mu)\bigr)}{\Psi(\mu)} =
  \frac{1}{2} \int_{\R^6} \rho(\bfr_1) \rho(\bfr_2)
  \left( \frac{1}{r_{12}} -w(r_{12},\mu) \right).
\end{equation}

Separating the Hartree part from $F[\rho]$ was already proposed by Hohenberg and
Kohn~\cite{HohKoh-PR-64}, and it is also the dominant part in $\bar{E}(\mu)$.
 Most molecular codes use Gaussian one-particle basis functions, so $\rho$ is a linear
combination of Gaussian functions.
We consider a generic term,
\begin{equation}
  \int_{\R^6} e^{- \alpha r_1^2} e^{- \alpha r_2^2}
  \left( \frac{1}{r_{12}} -w(r_{12},\mu) \right).
\end{equation}
This integral is easily computed, e.g., by using Fourier transforms, and one obtains an
expression that is proportional to the form of the basis function $\tilde \chi^{\ }_j$ 
in~\Eqref{eq:basis}, after arbitrarily relating $1/\alpha^2$ to the basis index $j$.

\begin{figure}[t]
  \centering
  \includegraphics[viewport=140 330 450 665,clip=]{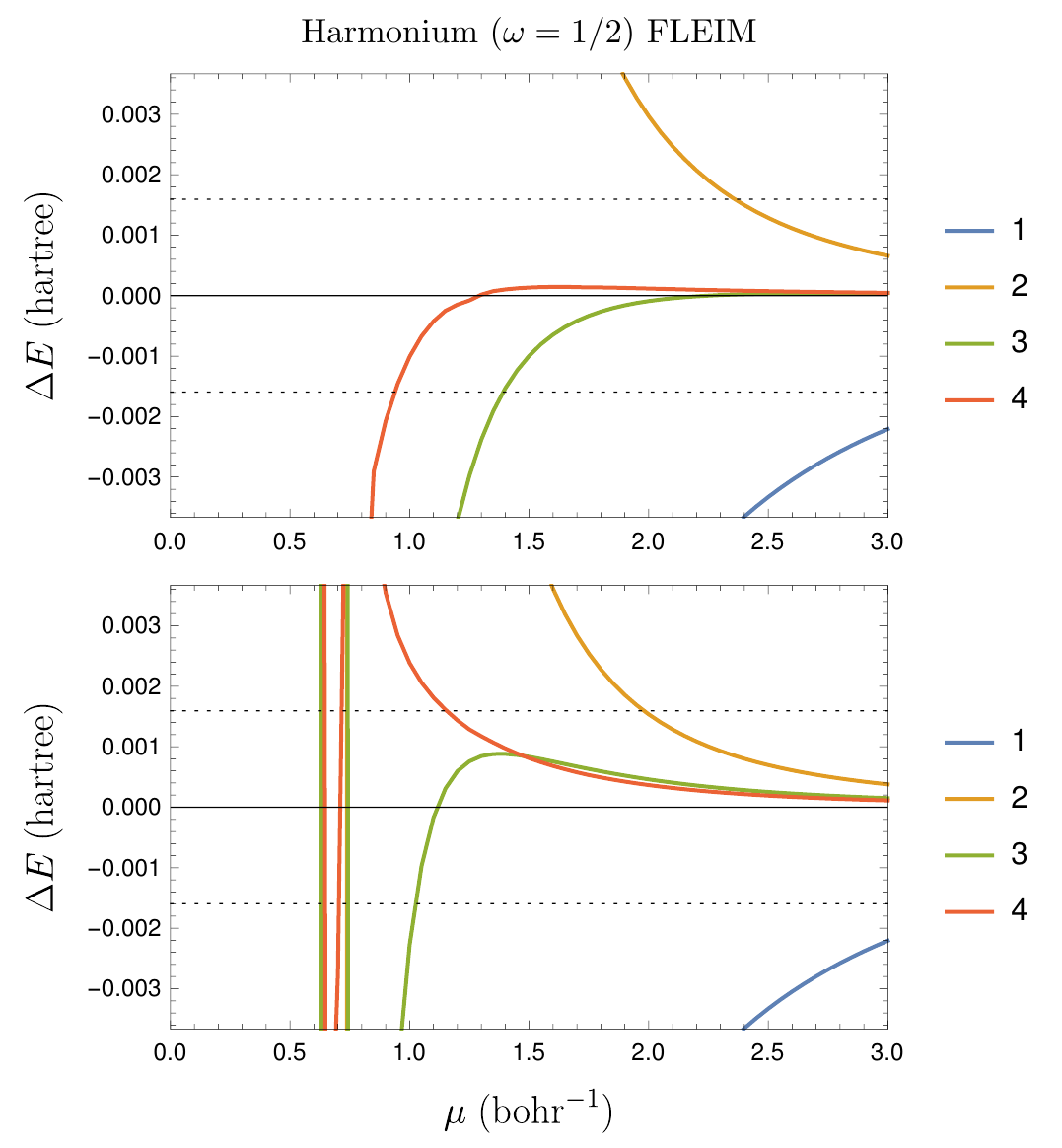}
  \caption{\label{fig:was10-11}%
Errors for harmonium using FLEIM with up to four points (1: blue curve, 2:
brown curve, 3: green curve, 4: red curve), using the basis sets given in
(\ref{eq:basis-decay}) (top) and (\ref{eq:basis-alt}) (bottom).  
The abscissa represents the biggest~\( \mu \) allowed for use in the FLEIM
algorithm.} 
\end{figure}

Let us now consider another basis set, constructed from the requirement that the functions
decay as $\mu^{-2}$ and are finite at the origin.
\begin{equation}
  \label{eq:basis-decay}
  \tilde \chi^{\ }_{1,j}(\mu)=\left(1+ (j \, \mu)^2 \right)^{-1}
\end{equation}
where we choose again $j=1,2,\dotsc, 10$.
The results are only slightly worse than those obtained when using the basis set
of~\Eqref{eq:basis}, compare Fig.~\ref{fig:was2-4}(top) and Fig.~\ref{fig:was10-11}(top).

Let us consider
\begin{equation}
  \label{eq:basis-alt}
  \tilde \chi^{\ }_{2,j}(\mu)=a_j \mu \left(1+ (a_j \, \mu)^3 \right)^{-1}
\end{equation}
where we choose $a_j=2^{j/2}$.
The result for harmonium is shown in Fig.~\ref{fig:was10-11}(bottom).
We note a slight improvement of the results for the 2- and 3-point approximations, as well
as a change of sign of the error in the 4-point approximation.
This direction of investigation deserves to be pursued.

\section{The empirical interpolation method}\label{app:EIM}
\enlargethispage{0.2\baselineskip}
The empirical interpolation method is a model-order reduction method introduced
in~\cite{barrault_empirical_2004} as a way to efficiently find a reduced basis and
approximate one particular function within a manifold of parameter dependent functions.
The points at which to do the interpolation are referred as magic
points~\cite{maday_general_2009}.

For a family of basis functions~\( \chi^{\ }_{i} \), \( i \in I \) with discrete
points~\( \mu_{j} \), \( j \in J \) chosen on a regular grid close to zero (see Appendix
\ref{667YU} for an analysis of the influence of the choice of the grid), we want to find a
family of \( K \)~functions and interpolation points with which to interpolate a test
function~\( f\in \mathrm{Span}\{\chi^{\ }_{i}, i \in I\} \).

\subsection{Algorithm for EIM}
We assume that we have chosen some cost function $\mathcal {C}$, e.g., a
norm~\( \mathcal{C}[\cdot] \equiv \norm{\cdot} \), or, if we are, as in this contribution, only
interested in correctly approximating the value for~\( \mu \equiv \infty \), we choose the
cost function as the absolute value at that extrapolation point
$\mathcal{C}[\varphi] = | \varphi(\infty)|=\lim_{\mu\rightarrow \infty} |\varphi(\mu)|$.

First, select one of the basis functions. We can choose to add the constant
function~\( \chi^{\ }_{0} \),
\begin{equation}
  \tilde \chi^{\ }_{0} \coloneqq \chi^{\ }_{0}.
\end{equation}
We then select the first interpolation point as the largest admissible~\( \mu \) available,
\begin{equation}
  \tilde \mu_{0} \coloneqq \max_{j \in J} \mu_{j}.
\end{equation}
We can then define the first normalized interpolation function as
\begin{equation}
  q_{0} \coloneqq \frac{\tilde \chi^{\ }_{0}}{\tilde \chi^{\ }_{0}(\tilde \mu_{0})}.
\end{equation}
We can then create the first approximation with an interpolation scheme, for instance
Lagrangian interpolation as
\begin{equation}
  \mathcal{I}_{0}[f] \coloneqq f(\tilde \mu_{0})q_{0}.
\end{equation}

We now assume to have chosen \( K-1 \)~functions \( \tilde \chi^{\ }_{k} \), normalized functions
\( q_{k} \).
Let us also assume that we have selected $K-1$
interpolation points \( \tilde \mu_{k} \), \( k = 0, \dotsc, K-2 \).
We define the \( \mathcal{I}_{K-2} \) Lagrangian
interpolation function as
\begin{equation}
  \mathcal{I}_{K-2}[f] \coloneqq \sum_{k=0}^{K-2} \beta_{k}q_{k},
\end{equation}
where the coefficients~\( \beta_{k} \) are determined by solving the system
\begin{equation}
  \sum_{k=0}^{K-2} \beta_{k}q_{k}(\tilde \mu_{\ell}) = f(\tilde \mu_{\ell})
  \quad\text{for\ }\ell = 0, \dotsc, K-2.
\end{equation}

If we denote \( \tilde I \) as the set of indices of remaining basis functions and
\( \tilde J \) as the set of indices of remaining interpolation points, we choose the next
function as
\begin{equation}
  \label{eq:eim_chi}
  \tilde \chi^{\ }_{K-1} \coloneqq \argmax_{\chi^{\ }_{i},\, i \in \tilde I}
  \bigl\{ \mathcal {C} \bigl[\chi^{\ }_{i} - \mathcal{I}_{K-2}[\chi^{\ }_{i}]\bigr]\bigr\},
\end{equation}
and the next interpolation point as
\begin{equation}
  \label{eq:eim_mu}
  \tilde \mu_{K-1} \coloneqq \argmax_{\mu_{j},\, j \in \tilde J}
  \bigl\{\abs*{\tilde \chi^{\ }_{K-1}(\mu_{j}) - \mathcal{I}_{K-2}[\tilde \chi^{\ }_{K-1}](\mu_{j})}\bigr\}.
\end{equation}
We can then define the \( K \)th normalised interpolation function as
\begin{equation}
  q_{K-1} \coloneqq \frac{\tilde \chi^{\ }_{K-1} - \mathcal{I}_{K-2}[\tilde \chi^{\ }_{K-1}]}{\tilde \chi^{\ }_{K-1}(\tilde \mu_{K-1}) - \mathcal{I}_{K-2}[\tilde \chi^{\ }_{K-1}](\tilde \mu_{K})}.
\end{equation}
This system is represented by a lower triangular matrix with ones on the diagonal, and hence
has a unique solution.
The algorithm ends when the desired target accuracy is reached.

\subsection{The forward looking empirical interpolation method (FLEIM)}
To better adapt the method for extrapolation, we propose a double loop alternative: Instead of
selecting sequentially first for a new basis function and then a new interpolation point ---
Eqs.~\eqref{eq:eim_chi} and~\eqref{eq:eim_mu} --- we select the best pair

\begin{equation}
  (\tilde \chi^{\ }_{K-1}, \tilde \mu_{K-1})
  \coloneqq \argmax_{\chi^{\ }_{i}, \, i\in\tilde I}
  \argmin_{\mu_{j},\, j\in\tilde J} \bigl\{ \mathcal {C} \bigl[\chi^{\ }_{i}(\mu_{j}) - \mathcal{I}_{K-2}[\chi^{\ }_{i}](\mu_{j})\bigr]\bigr\}.
\end{equation}

\section{Numerical details of the calculations}\label{app:details-calc}
\subsection{Testing EIM and FLEIM with $E(\mu) = 1 + \chi^{\ }_j (\mu)$}
\label{app:details-calc1}
\begin{figure}[p]\centering
  \begin{sideways}
  \includegraphics[width=480pt]{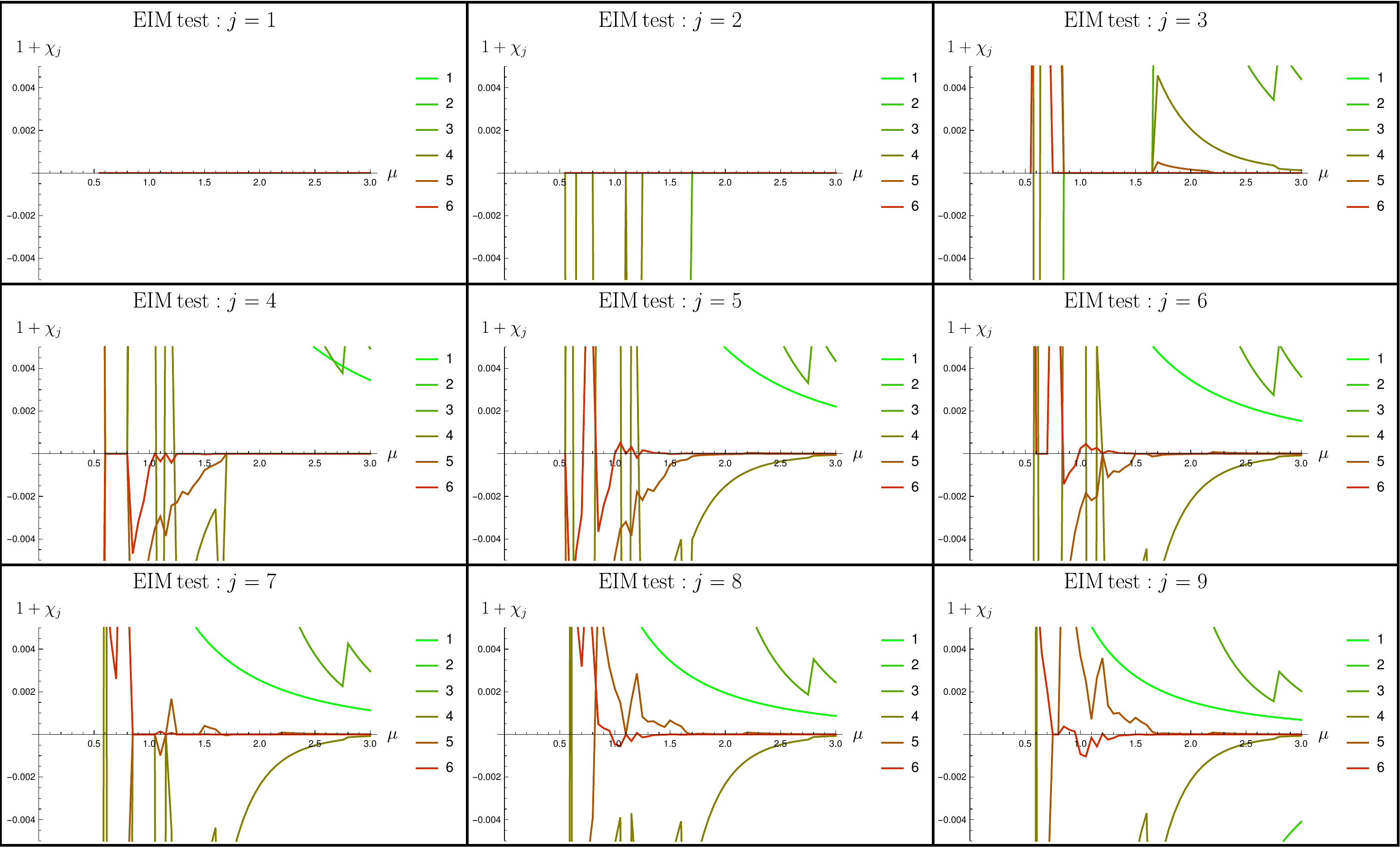}    
  \end{sideways}
  \caption{Test of EIM using model
    $E(\mu) = 1 + \chi^{\ }_j (\mu)$.\label{fig:test-EIM}}
\end{figure}

\begin{figure}[p]\centering
  \begin{sideways}
  \includegraphics[width=480pt]{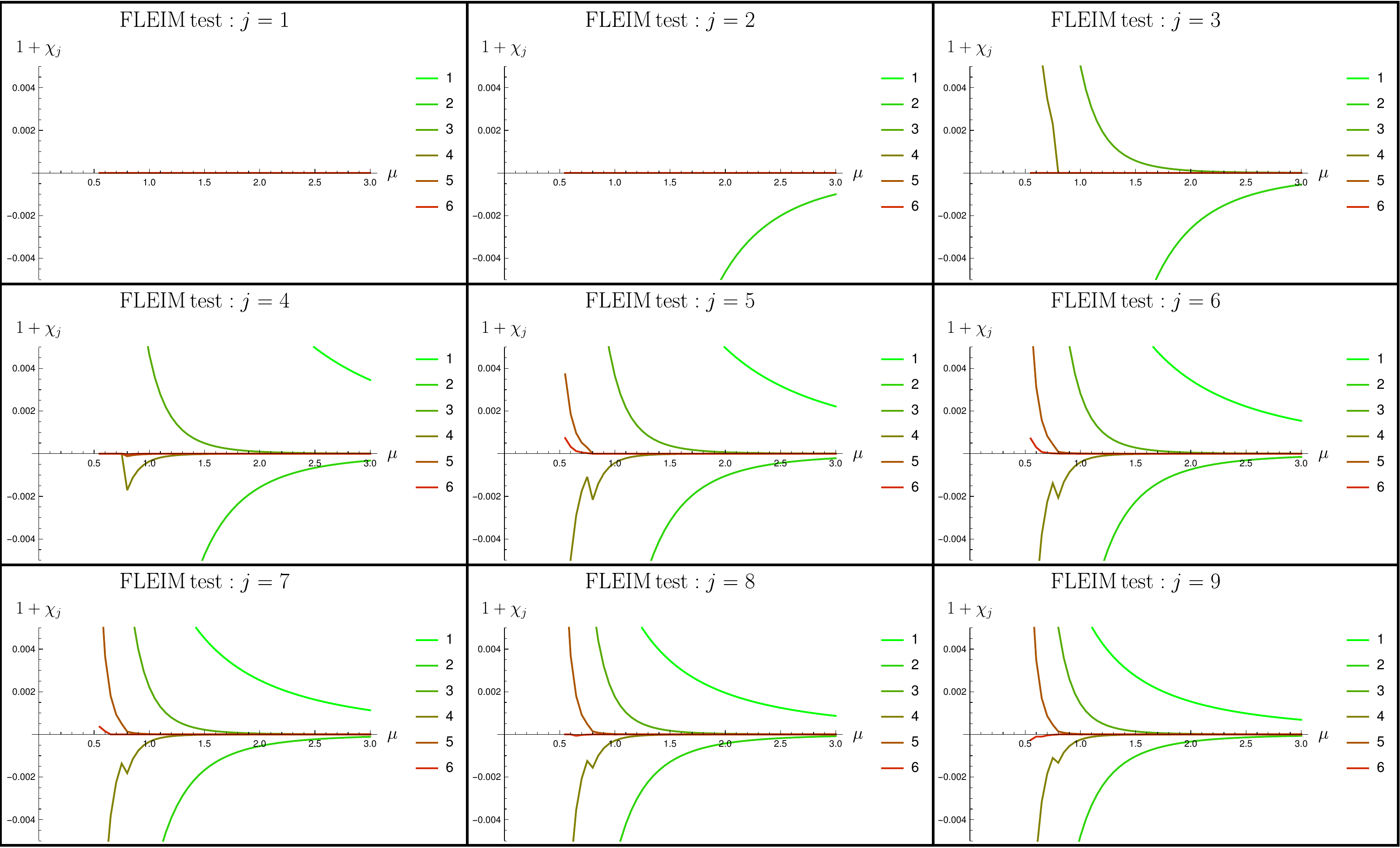}    
  \end{sideways}
  \caption{Test of FLEIM using model
    $E(\mu) = 1 + \chi^{\ }_j (\mu)$.\label{fig:test-FLEIM}}
\end{figure}
In this subsection, we compare on a simple function, the behaviour of EIM and FLEIM on a
analytic test function $E(\mu)$ behaves like $1+\chi^{\ }_j(\mu)$.
The results  for EIM are given in Fig.~\ref{fig:test-EIM}, and for FLEIM, in
Fig.~\ref{fig:test-FLEIM}.

First, we note that FLEIM provides better approximation together with more stable results
when $\mu$ varies.
Second, we note that the ``wall'' for $j>1$ at small $\mu$ is also imporved.
At large $\mu$, all $\chi^{\ }_j$ have the same decay at large $\mu$: this makes both methods fit
to work in this regime.

\subsection{Discretization for FLEIM}\label{667YU}
The interval between \num{0} and the value of $\mu$ under study was divided in \num{10}
equal intervals (producing \num{11} points).
FLEIM was used to select $K (\le 4) $ points and basis functions on which $E(\mu)$ was
calculated.

\begin{figure}[p]
  \centering
  \includegraphics[viewport=140 165 450 635,clip=]{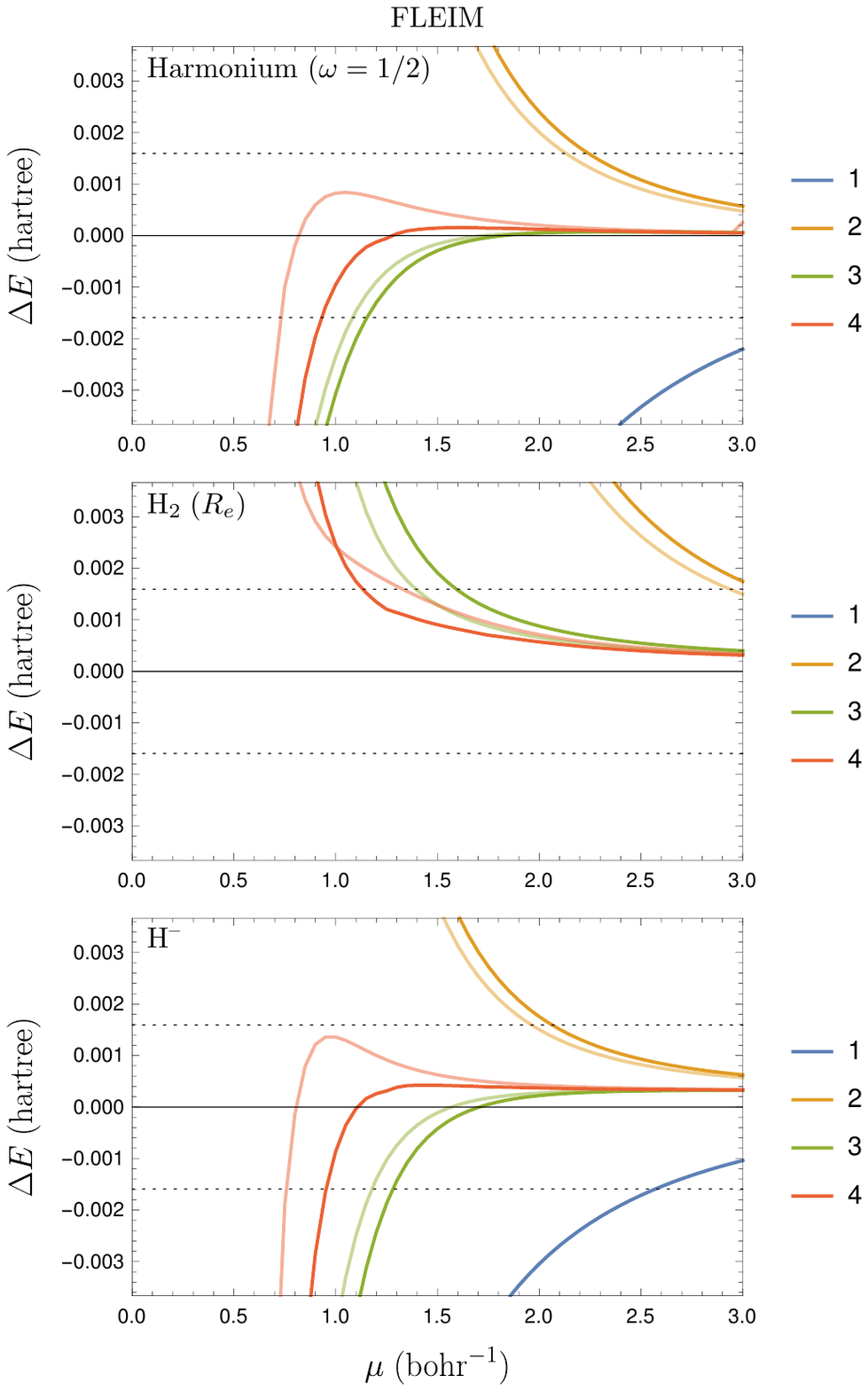}
  \caption{\label{fig:densergrid}%
    Results using a denser grid (lighter colors), for harmonium, \htwo{} at
    equilibrium distance, and \hminus{}; the curves obtained with the denser grid are shown
    in lighter colors.
    The error of the model without correction (blue curve) does not show up 
in the figure for the \htwo\ molecule because it is larger than the 
domain covered by the plot.}
\end{figure}

We now investigate the effect of changing the grid of $\mu$ and use, for comparison a fixed
finer grid of values of $\mu_1, \mu_2, \dotsc$ ranging from 0 to the value $\mu$ indicated
in the plots, with a step of \SI{0.01}{\per\bohr}, the FLEIM results are only slightly
changed, as shown in Fig.~\ref{fig:densergrid}.

\subsection{Systems}
Two-electron systems studied are
\begin{enumerate}
\item harmonium, having
  \begin{equation}
    \label{eq:v-harm}
    v(\bfr)=\frac{1}{2} \omega^2 r^2
  \end{equation}
  where for $\omega=1/2$ the exact energy is known ($E=\SI{2}{\hartree}$);
\item \hminus{} anion,
  \begin{equation}
    \label{eq:v-hm}
    v(\bfr)=-\frac{1}{r};
  \end{equation}
\item \htwo{} molecule,
  \begin{equation}
    \label{eq:v-h2}
    v(\bfr)=-\frac{1}{\abs{\bfr -\bfR_A}}-\frac{1}{\abs{\bfr -\bfR_B}}
  \end{equation}
  where the nuclei are in the equilibrium position, ${\bfR_A = -\bfR_B}$ with
  $\abs{\bfR_A}=\SI{0.7}{\bohr}$.
\end{enumerate}

\subsection{Obtaining the model energy}
In order to simplify the test of FLEIM, $E(\mu)$ was pre-calculated for a dense range of
values $\mu$ and interpolated.
The values for $E(\mu)$ were obtained with the program Molpro~\cite{Molpro-PROG-02}  for
\hminus{} and \htwo.
This program was also used for the density functional calculations.

For harmonium, it is possible to separate the variables in the Schr\"o\-dinger equation.
The center-of-mass equation can be solved exactly.
The equation in $r_{12}$ was solved by discretization on a grid of \num{e4}~points between
\num{0}~and~\SI{10}{\bohr}.

For \htwo{} the V5Z basis set of \cite{Dunning-JCP-89}
was used;
for \hminus\ the aug-V5Z basis set of \cite{Woon-Dunning-JCP-94}.
(The error in the energy of \hminus{} is too large if we do not augment the V5Z basis set
with a diffuse basis function.)
The aug-V5Z basis set was also used for the excited state of \htwo, as it has an important
contribution of ionic states (\hplus$\cdots$\hminus).
For \htwo{} the equilibrium distance of \SI{1.4}{\bohr} was chosen for the ground state.
For the excited state, the distance of \SI{4.2}{\bohr} was chosen. It is close to a minimum
of the potential energy curve. Furthermore, this value can be compared with the accurate
calculation of Ref.~\cite{NakNak-JCP-18}.

\section{Change of coordinates in harmonium}\label{app:change-to-cm-12}
Harmonium is characterized by the external potential given by~\Eqref{eq:v-harm}.
For two particles the variables $\bfr_1, \bfr_2$ can be changed to those corresponding to
the center of mass and the distance between particles,
\begin{eqnarray}
  \label{eq:eqs-r}
  \bfR & = & \frac{1}{2} \left(\bfr_1 + \bfr_2 \right),\nonumber \\
  \bfr_{12} & = &  \left(\bfr_1 - \bfr_2 \right)
\end{eqnarray}
yielding for the potential energy
\begin{equation}
  \label{eq:vcm+v12}
  \frac{1}{2} \omega^2 \left(\bfr_1^1 + \bfr_2^2 \right) =
  \omega^2 \left( \bfR^2  + \frac{1}{4}\bfr_{12}^2 \right)
\end{equation}
The transformation of variables can be done also for the kinetic energy, and makes the model
Schr\"odinger equation separable into a part (in $\bfR$) that is independent of the model,
and one (in $\bfr_{12}$) that through \ref{eq:werf} depends on the model ($\mu$).

Note that $\bra{\Psi(\mu)} r_{12}^2 \ket{\Psi(\mu)}$ is also measuring the error due to the
change of density with $\mu$, for harmonium

\begin{equation}
H(\mu)  =  T + \frac{1}{2} \omega^2 (r_1^2 +r_2^2) +\frac{\erf(\mu r_{12})}{r_{12}}
\end{equation}
Indeed, by using the Hellmann--Feynman theorem,
\begin{eqnarray}
  \frac{\partial }{\partial \omega^2} \bra{\Psi(\mu)} H(\mu) \ket{\Psi(\mu)}
  & = &
  \bra{\Psi(\mu)} \frac{1}{2} (r_1^2 +r_2^2) \ket{\Psi(\mu)}\nonumber \\
  & = & \bra{\Psi(\mu)} ( R^2 +\frac{1}{4} r_{12}^2) \ket{\Psi(\mu)}
\end{eqnarray}
where $\vecfont{R}=(\bfr_1+\bfr_2)/2$.
As it is possible to separate variables $\bfR$ and $\bfr_{12}$ in the Schr\"odinger
equation,
\begin{equation}
  \Psi(\mu)= \Psi(\bfR, \bfr_{12},\mu) = \Phi(\bfR) \phi(\bfr_{12},\mu)
\end{equation}
$\Phi$ and $\phi$ both normalized to one, and
\begin{equation}
  \bra{\Psi(\mu)} \frac{1}{2} (r_1^2 +r_2^2) \ket{\Psi(\mu)}
  = \int_{\R^3} \rho(\bfr,\mu) r^2 \,\D\bfr
\end{equation}
where $\rho(\bfr,\mu)$ is the density of the model system, we have
\begin{equation}
  \frac{1}{4} \bra{\Psi(\mu)} r_{12}^2 \ket{\Psi(\mu)} =
  \int_{\R^3} \rho(\bfr,\mu)\, r^2 \,\D\bfr
  - \int_{\R^3} |\Phi(\bfR)|^2 R^2 \,\D\bfR
\end{equation}
Note that the change of $\langle r_{12}^2 \rangle$ with $\mu$ is only due to
the change of the density with $\mu$.
Thus, the difference between
$\frac{1}{4} \bra{\Psi(\mu)} r_{12}^2 \ket{\Psi(\mu)}$ and
$\frac{1}{4} \bra{\Psi} r_{12}^2 \ket{\Psi}$
also indicates how much the density is affected by the model.

\section*{Funding}
Part of this work was supported by the French ``Investissements d'Avenir'' program, project
ISITE-BFC (contract ANR15-IDEX-0003) (\'EP).
This work has also received funding from the European Research Council (ERC) under the
European Union’s Horizon 2020 research and innovation program (grant agreement No 810367--project EMC2) (YM \& AS).

\end{document}